\documentclass[aps,pra,twocolumn]{revtex4}
\usepackage{epsfig}
\usepackage{amsmath,amsthm,amssymb}

\newcommand{\aone}{\vec{a}_1}
\newcommand{\atwo}{\vec{a}_2}
\newcommand{\athree}{\vec{a}_3}
\newcommand{\cone}{\vec{c}_1}

\begin{document}

\title{On the study of jamming percolation}

\author{M. Jeng\footnote{mjeng@physics.syr.edu} and  
J. M. Schwarz\footnote{jschwarz@physics.syr.edu}}


\affiliation{
Physics Department, 
Syracuse University,
Syracuse, New York 13244, USA}


\begin{abstract}
We investigate kinetically constrained models of glassy transitions, and determine which model characteristics are crucial in allowing a rigorous proof that such models have discontinuous transitions with faster than power law diverging length and time scales. The models we investigate have constraints similar to that of the knights model, introduced by Toninelli, Biroli, and Fisher (TBF), but differing neighbor relations. We find that such knights-like models, otherwise known as models of jamming percolation, need a ``No Parallel Crossing'' rule for the TBF proof of a glassy transition to be valid. 
Furthermore, most knight-like models fail a ``No Perpendicular Crossing'' requirement, and thus need modification to be made rigorous.
We also show how the ``No Parallel Crossing'' requirement can be used to evaluate the provable glassiness of other correlated percolation models, by looking at models with more stable directions than the knights model. Finally, we show that the TBF proof does not generalize in any straightforward fashion for three-dimensional versions of the knights-like models.
\end{abstract}

\maketitle

\section{Introduction}

The puzzle of glass-forming systems has remained sufficiently elusive 
over the years such that even the puzzle pieces themselves have changed shape.  For example, the 
puzzle piece of the lack of a growing lengthscale may now have to be 
modified since a growing lengthscale can perhaps be extracted from a 
higher-order correlation function~\cite{glassexp1}. A less recent change in puzzle pieces 
is the distinction by Angell between fragile and strong glasses, where certain effects 
are more dramatic in fragile glasses than in strong ones~\cite{glassexp2}. 
One piece of the puzzle that has 
remained constant over the years, however, is the dramatic slowing 
down of the dynamics of the 
particles near the glass transition.
More precisely, a supercooled liquid's viscosity can 
increase by fourteen orders 
of magnitude as the temperature is 
decreased near a ``working'' definition of 
the glass temperature~\cite{glassexp2}.

Two main phenomenological models for this dynamical 
slowing down have emerged over the years: mode-coupling theory and 
kinetically constrained models.    We will not focus on mode-coupling theory here and 
simply refer the reader to several excellent reviews~\cite{gotze,kob,das}.
As for the second phenomenological approach, kinetically constrained models, 
the goal is to understand whether glassy dynamics can be understood as
arising from steric constraints on the particles alone~\cite{KCM.Review}. 
One of the simplest such examples is the Kob-Andersen 
model~\cite{KobAndersen}.  It is  motivated by the caging of particles, ultimately  
observed in larger scale systems such as colloidal glasses~\cite{Colloid.Caging.1,Colloid.Caging.2}.
The Kob-Andersen model is a hard-core 
lattice gas model, but with the constraint that a
particle can only hop
to an adjacent site if and only if it has less than a certain number of
neighbors, $m$, both before and after the move.
Early simulations of the Kob-Andersen model 
on the hypercubic lattice for relevant values of $m$ appeared to find
a dynamical phase transition at a nontrivial critical
density~\cite{KobAndersen}.
However,
subsequent mathematically rigorous results  
found that the phase transition does not occur until the 
fully packed state~\cite{TBF.KA.PRL,TBF.KA}.
This corresponds to a zero-temperature glass transition.
TBF proved this by showing that at any monomer density,
there were mobile cores that could diffuse at sufficiently
long time scales.

While the hypercubic version of the Kob-Andersen model does not 
exhibit a finite-temperature glass transition, the mean field version 
does~\cite{TBF.KA.PRL,TBF.KA}.   Since it is still up 
for debate whether or not mean field is relevant 
for physical systems, one can ask whether or not there exists 
another finite-dimensional, kinetically constrained model that exhibits a 
finite temperature glass transition.  While the Kob-Andersen (and the 
Fredrickson-Andersen~\cite{fredrickson.andersen}) models,
are elegant in their simplicity, there are indeed
two more involved 
kinetically constrained models in two dimensions that can be proven to exhibit a 
finite-temperature glass 
transition~\cite{TBF,TB1,TB2}.  These two models have been 
dubbed the spiral model~\cite{Reply} and the sandwich 
model~\cite{Comment}. Both 
models exhibit an unusual phase transition in that 
the fraction of frozen particles jumps discontinuously at the 
transition, typical of a first-order phase transition. 
However, as the transition is approached from below, there 
exists a crossover lengthscale that diverges faster 
than a power law in $T-T_g$. The crossover length, $\Gamma$, 
distinguishes between squares
of size $L<<\Gamma$, which are likely to contain a frozen
cluster, and squares of size $L>>\Gamma$, for which the probability of containing a frozen cluster is exponentially 
unlikely.  Given this combination of a discontinuity in the
fraction of frozen particles, and a faster than power law diverging length scale, the transition has unique characteristics~\cite{unique}.

Models such as the spiral and sandwich models
are proof in principle that further exploration 
of kinetically constrained models in finite-dimensions may be 
fruitful, in particular, because should an ideal glass transition 
exist, it may indeed be of unusual character. More specifically, the
Edwards-Anderson order parameter should be discontinuous at
the transition, yet accompanied by rapidly diverging time 
scales~\cite{GlassGeneral,SpinFacilitated.Review}.

Kinetically constrained models can be related to models of
correlated percolation, which are models in which each site
is initially occupied with an independent probability $\rho$,
as in normal percolation,
but then correlations are induced through some culling
rule---that is, sites can only be occupied if certain
conditions on their neighboring sites are met. 
The two types of models are related, with the
immobile particles of the kinetically constrained models
corresponding to the stable occupied particles of the
correlated percolation models.
The simplest model of correlated percolation is 
$k$-core percolation, in which an occupied site must have at
least $k$ occupied neighbors~\cite{kcore1,kcore2}. Occupied sites that do not
satisfy this stability requirement are removed, and this
condition is applied repeatedly, until all remaining sites
are stable. This model can be mapped to the Fredrickson-Andersen 
model~\cite{KCM.Review,sellitto}.

One might think that kinetically constrained models and 
correlated percolation systems would be easy to numerically
simulate, and that construction of rigorous proofs would be nothing more than an
interesting problem for mathematicians. However, while
simulating these systems is easy, extracting their properties
in the infinite system size limit is much harder.
Just as with the Kob-Andersen model, initial numerical
simulations of $k$-core percolation 
for certain values of $k$
found evidence of
first-order phase transitions at nontrivial critical
densities, and of second-order phase transitions in a
different universality class than normal 
percolation~\cite{kcore.odd.old.1,kcore.odd.old.2,kcore.odd.old.3}.
However, subsequent mathematically rigorous analyses found that the 
critical point is $\rho_c=1$
for those $k$~\cite{kcore.rigorous.1,kcore.rigorous.2,kcore.rigorous.3}. Because the critical point only approaches unity very slowly
in the limit of infinite system size, the simulations on
finite-size systems were misled as to the location of the
true critical point, thereby highlighting the importance of rigorous 
results for these models as well.  
For a review of $k$-core percolation, see
Ref.~\cite{kcore.review}.

There is another finite-dimensional system exhibiting an unusual transition, at least 
numerically.  It is the jamming transition in repulsive soft spheres. 
Numerical simulations of soft spheres show
a critical point at which the average coordination number
jumps discontinuously to a universal, isostatic 
value. But quantities such as the shear modulus and the deviation of the
average coordination number from its isostatic value show
a nontrivial power law behavior in the vicinity of the critical 
point~\cite{SphereSims1,SphereSims2,SphereSims3}.  Recent experiments 
on two-dimensional photoelastic beads support this notion of a mixed 
transition~\cite{behringer}. 

Interestingly, 
it has been conjectured that the physics of granular systems, 
colloidal systems, and glassy systems are of a similar 
character~\cite{JammingPhaseDiagram}.  
The mean field results of not one but several correlation percolation models corresponding 
to glassy physics support this notion quantitatively.  Furthermore, 
experimental evidence of caging in another two-dimensional granular system 
also supports
this notion~\cite{Granular.Caging}. The question 
of finite-dimensional glassy models being quantitatively similar to 
the repulsive soft sphere system is still being investigated.  Certainly 
the spiral and sandwich models show that, qualitatively, one can have 
a glassy system exhibiting an unusual 
finite-dimensional transition. However, they do
not appear to be in the same class as the jamming system, since 
the order parameter exponent is unity just above the transition 
in the jamming percolation models, but is one-half in the jamming 
system.

To explore the possible link between jamming and glassy systems in terms 
of a finite-temperature glass transition, TBF initially introduced the knights model, a model of
correlated percolation similar to the spiral 
model~\cite{TBF}, and called it a model of jamming percolation. In fact, 
the spiral, knights, and sandwich models are all models
of jamming 
percolation.
In this paper, we expand on our earlier work
(Ref.~\cite{Comment}),
in which we introduced the sandwich model,
by presenting the details of the proof of an unusual
transition in this model.  This proof is
based on modifying 
the proof developed by TBF in Refs.~\cite{TBF,TB1,TB2} for 
the spiral model (although originally misapplied to the knights model).    
We also introduce further generalizations 
of models of jamming percolation to demonstrate that the phenomenon 
of a finite-dimensional transition is indeed somewhat generic. 
In doing so, we show that the TBF proof only gives a rigorous
derivation of these novel properties if a ``No parallel
crossing'' rule holds. This rule says that two
similarly-oriented directed percolation chains
cannot cross without having sites in common.
The effect of this rule is that 
one directed percolation-like process cannot be used
to locally stabilize the other.
For models such as the sandwich model,
which satisfy this ``No parallel crossing'' rule,
but fail a ``No perpendicular crossing'' rule,
the TBF proof works only with some modification.

The methods described in this paper can be used to
understand for which correlated percolation models a proof
along the lines of the TBF proof can be used to show a
glassy transition, and for which models they cannot. Given that there 
are two very detailed papers on the TBF proof~\cite{TB1,TB2}, 
we will refer to 
them quite often, as opposed to making this paper self-contained. 
Finally, we will discuss connections between jamming percolation 
and force-balance percolation, another correlated percolation 
model inspired by granular systems, where numerical evidence 
points toward an unusual transition in finite 
dimensions~\cite{fbperc1,fbperc2}.


\section{The class of models}

\begin{figure}[tb]
\epsfig{figure=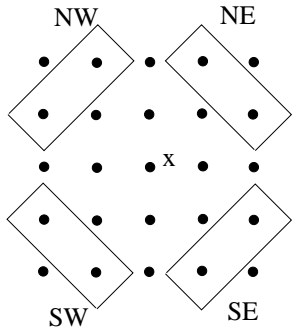,width=1.5in}
\caption{The knights model neighbors.}
\label{fig:knights}
\end{figure}

\begin{figure}[tb]
\epsfig{figure=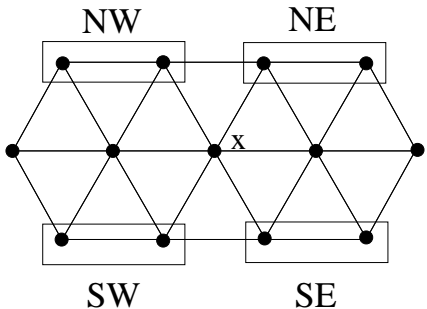,width=2.2in}
\caption{The sandwich model neighbors}
\label{fig:sandwichModel}
\end{figure}

\begin{figure}[tb]
\epsfig{figure=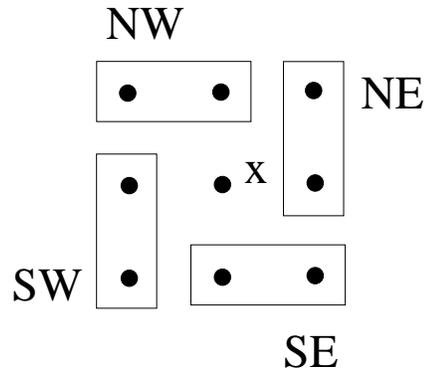,width=2.2in}
\caption{The spiral model neighbors}
\label{fig:spiralModel}
\end{figure}

We consider a class of models that generalizes the knights
model, the earliest of the jamming percolation models.  The 
class of models is 
defined
on the two-dimensional square 
lattice.
Initially, each site is occupied with 
an independent probability $\rho$.  
Each site has four neighboring sets, and each set contains two
sites.
The four sets are labelled as the northeast, northwest,
southeast, and southwest neighboring sets.
(The sites in those sets only lie in precisely those compass
directions for the knights model, but we continue to label
the four sets in this manner for all our correlated
percolation models
until section~\ref{sec:The implications of Property B}.)
To be stable, an occupied site must 
either have (1) at least one northeast neighbor and at 
least one southwest neighbor, or
(2) at least one northwest neighbor and at least one southeast 
neighbor.  All other occupied sites are unstable, and are 
vacated.
This culling process is then repeatedly applied---sites 
that were previously stable may become unstable by earlier
cullings---until all remaining sites are stable. 
The neighboring sets of the original knights model
introduced by TBF~\cite{TBF} are shown in 
Fig.~\ref{fig:knights}.
Fig.~\ref{fig:sandwichModel} shows the 
neighboring sets in the sandwich model, which we
introduced in~\cite{Comment}, and 
Fig.~\ref{fig:spiralModel} shows the neighboring sets in
the spiral model, which TBF introduced in~\cite{Reply}.

In these models, stable sites must either be part of 
a chain running from the northeast to the southwest, or
part of a chain running from the northwest to the
southeast. In the final configuration, all sites must be
stable, so any chain must either continue forever (to the
boundary), or terminate in a chain of the
other type (this latter case is called a ``T-junction'').
Any sites left after the culling procedure
must thus be connected by an infinite series of chains (or, in a 
finite system, connected
by a series of chains to the distant boundary), so
asking for the critical probability
at which an infinite
cluster first appears
is the same as asking for the minimum probability at 
which
some sites remain unculled in the infinite size limit.

\begin{figure}[tb]
\epsfig{figure=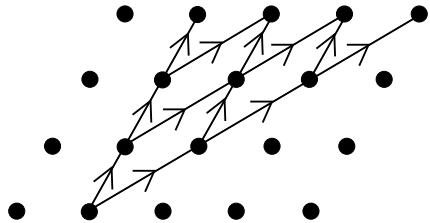,width=2.2in}
\caption{A sublattice in the sandwich model.}
\label{fig:Sublattice.sandwich}
\end{figure}

If sites could only be stable by having northeast and
southwest
neighbors, then every stable site would
be part of a northeast-southwest chain on a 
particular sublattice---the sublattice extending to the
northeast of the site for the sandwich model is shown
in figure~\ref{fig:Sublattice.sandwich}. Chains on this
sublattice are isomorphic to infinite chains in directed percolation,
which has a well-studied second-order phase transition~\cite{DP.review}.
However, in these models there is an additional mechanism by
which sites can be made stable---that is,
by having northwest and
southeast neighbors. 
Adding an extra way for a sites to be stable
can only possibly depress the critical
probability, so we immediately see
that for these models
$\rho_c\leq \rho_c^{\rm DP}$.

It will turn out to be useful to divide these models into
classes based on two properties.
We define a model as having ``No Parallel Crossing''
property if whenever
two northeast-southwest chains
(or two northwest-southeast chains) intersect, they must
have sites in common---we abbreviate this a
``Property A.''
And we define a model as having a ``No Perpendicular
Crossing'' property, or Property B, if
whenever a northeast-southwest and northwest-southeast 
chain intersect, they must
have sites in common.
Table~\ref{table:properties} shows
which properties each of the 
three models possesses:

\begin{eqnarray}
\nonumber
\begin{tabular}{l|c|c}
Model & Property A & Property B \\
\hline
knights & No & No \\
sandwich & Yes & No \\
spiral & Yes & Yes 
\end{tabular}
\label{table:properties}
\end{eqnarray}

\bigskip

\begin{figure}[tb]
\epsfig{figure=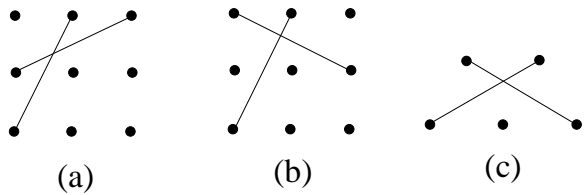,width=3.0in}
\caption{(a) Failure of Property A (``No Parallel
Crossings'') in the knights model.
(b) Failure of Property B (``No Perpendicular Crossings'')
in the knights model.
(c) Failure of Property B 
in the sandwich model.}
\label{fig:Crossings}
\end{figure}

\noindent Examples of where the properties fail for each
model are shown in figure~\ref{fig:Crossings}.

Our analysis here is based to a large extent on the claimed  
TBF proof of a glassy transition for the knights model.
Their proof consisted of two parts.
First, they claimed to show that the critical
point of the knights model is exactly the same as that for
directed percolation. Second, once this was done, they were
able to use well-known results on directed percolation
(assuming a well-tested conjecture about anisotropic
rescaling in directed percolation) to show that this model
has a glassy transition---specifically, they were able to
find structures with a finite density at the critical point
of directed percolation,
and to show that just below this critical point, the
crossover length and culling times diverged.
A short version of their proof appeared in
Refs.~\cite{TBF}, with more detailed explanations in
Ref.~\cite{TB1} and~\cite{TB2}.

Our analysis of this more general class of models
shows that the TBF proof that
$\rho_c=\rho_c^{\rm DP}$ is only valid for models
satisfying property A; so it works for the sandwich and
spiral models, but fails for the knights model.
The second part of their proof, showing a glassy
transition, implicitly assumes property B.  The spiral model exhibits property 
B and hence the TBF method of proof carries through~\cite{correction}.  
However, we show that the proof can be modified to work for
models that fail to have property B.
The spiral {\it and } the sandwich models thus have provably glassy
transitions, while the knights model does not.


\section{Identifying the critical point}
\label{sec:critical.point}

\begin{figure}[tb]
\epsfig{figure=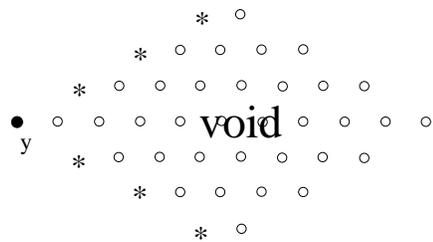,width=2.2in}
\caption{A void in the sandwich model. If the site $y$ is
unstable, its removal will cause the culling of all the
sites marked with stars. 
The corresponding void for the 
spiral model is a diamond, regular in shape.}
\label{fig:sandwich.Void}
\end{figure}

We sketch the TBF proof that 
$\rho_c^{\rm knights}=\rho_c^{\rm DP}$,
which will let us
understand 
why property A is sufficient, and most likely necessary
to the result.
The key to the TBF proof is to show that voids (clusters of
empty sites) of particular shapes have a finite probability
of growing forever.
For example, for the
diamond-shaped void in the sandwich model, shown
in figure~\ref{fig:sandwich.Void}, 
if the key site labelled $y$ is vacant, it will trigger the
removal of all the sites marked with stars, increasing the
void size by one.
(The corresponding void for the knights model
appears as
figure 1c of~\cite{TBF}.)
If this process repeats forever, with such key sites 
repeatedly removed, this void will grow to
infinity.

The TBF proof for the knights model
is based on the claim that the key sites of the
knights model, located at the corners of octagonal voids
of size $L$,
can only be stable if part of a directed
percolation chain of $\mathcal{O}(L)$.
Once this claim is granted, the rest of the 
proof is straightforward.
For $\rho < \rho_c^{\rm DP}$, such long chains are
exponentially suppressed, and thus for a large void, the
vertices at the corners of the void are exponentially likely to be
culled. Summing up the relevant probabilities, this results
in a finite probability that the void will grow to infinity.
For an infinite lattice, it is thus certain that there will be at
least one void that grows to infinity, showing that all
sites are culled below $\rho_c^{\rm DP}$.
Since $\rho_c^{\rm knights}\leq \rho_c^{\rm DP}$,
this is supposed to show that
$\rho_c^{\rm knights}=\rho_c^{\rm DP}$.

\begin{figure}[tb]
\epsfig{figure=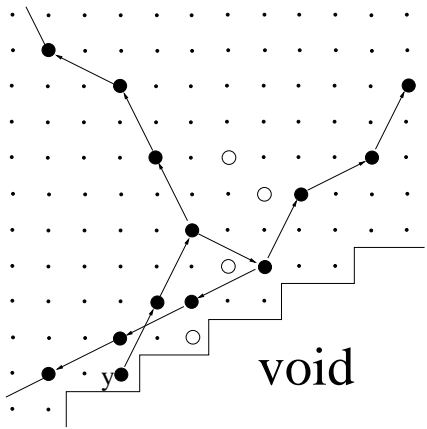,width=2.2in}
\caption{A counterexample to the claim in
Refs.~\cite{TBF,TB1,TB2} that for $y$ to be stable it
must be part of a long, uninterrupted chain to the northeast.
Large solid circles are occupied
sites, and large empty circles are vacant sites. 
All sites in the void are vacant.
All other sites
can be either occupied or unoccupied.
Arrows are drawn from each occupied site to indicate the
neighboring sites that make it stable.}
\label{fig:Counterexample}
\end{figure}

The claim that sites at the corners of voids can only be 
stable if part of a long chain of $\mathcal{O}(L)$
is only valid for models with property A.
A counterexample to this claim for the knights
model can be seen in
Fig.~\ref{fig:Counterexample}.
This counterexample also shows why
property A is necessary and sufficient for this claim to 
be valid. To be stable, the site must be part of a
northeast-directed chain; pick the lowest northeast-directed
chain coming out of the corner site $y$. 
If that chain stops before
reaching length $\mathcal{O}(L)$, it must terminate in a 
northwest-southeast chain. Since the northeast chain
is not as long as a wall of the void, the new
southeast-directed chain will eventually hit the void
(thus resulting in the culling of all chains, and the corner
site $y$), unless it hits a T-junction.
That new T-junction results a second northeast-southwest
chain, which will eventually reach the first northeast
chain, as in figure~\ref{fig:Counterexample}.
For models with property A, the two chains will intersect,
contradicting the original assumption that we chose
the lowest northeast-directed chain out of $y$.
Thus, by contradiction, for models with property A, the 
first northeast chain must be $\mathcal{O}(L)$
for $y$ to be stable,
and it indeed follows that
$\rho_c=\rho_c^{\rm DP}$.

What about for models such as the knights model, that lack
property A?
Is it possible that despite this counterexample
to the claim,
$\rho_c^{\rm knights}$ is actually equal to $\rho_c^{\rm
DP}$,
for some other reason?
While we do not have a mathematically rigorous proof
that $\rho_c^{\rm knights}\neq \rho_c^{\rm DP}$, we present an argument
that the two are 
almost certainly unequal.
We present our arguments in the context of the knights model,
but they generalize to other models that lack property A.

\begin{figure}[tb]
\epsfig{figure=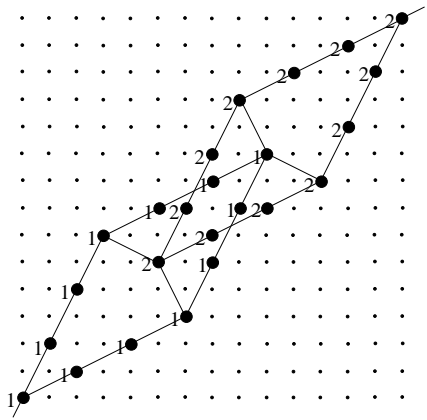,width=2.2in}
\caption{A substructure that depresses the critical point of
the knights model. Numbers indicate sites in the same
sublattice.}
\label{fig:Sub1to2}
\end{figure}

Consider the substructure in
Fig.~\ref{fig:Sub1to2}.
All sites in it are stable under the knights model culling
rules, except for the two sites at its ends, and those sites
will become stable if the substructure is attached 
between two
northeast-southwest chains. Furthermore, there is no
northeast-southwest chain 
internal to the substucture connecting the two ends.
The substructure is internally stabilized by
northwest-southeast links.

This means that the substructure can act as a ``rest stop.''
Northeast-southwest chains can have breaks in their paths,
connected by this substructure.
Just below $\rho_c^{\rm DP}$, long northeast-southwest
chains almost form an infinite structure. They are almost
linked, so a few extra connections, through these
substructures, should create an infinite cluster
even below $\rho_c^{\rm DP}$. So we expect that
$\rho_c^{\rm knights}<\rho_c^{\rm DP}$.

\begin{figure}[tb]
\epsfig{figure=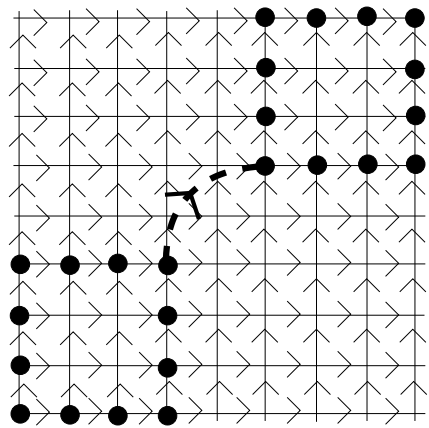,width=2.2in}
\caption{A configuration in ``jumping directed
percolation'' that receives an additional connection with
probability $s$.}
\label{fig:JumpPerc}
\end{figure}

We can make the argument more formal by considering
the following modification of the directed
percolation problem, which we call jumping directed
percolation. 
As 
with normal directed percolation,
we occupy sites on the square lattice
with probability $\rho$, and connect each site with
directed bonds to its neighbors to the north and east.
However, now we define an
additional way for sites to be connected. We divide the
lattice into blocks of size 9x9, and for each block, if the
two hollow squares of sites shown in Fig.~\ref{fig:JumpPerc} 
are occupied,
we with probability $s$ connect the two hollow squares 
with a directed bond from the southwest square to the
northeast square.
The critical point of this model is a
function of $s$: $\rho_c^{\rm Jump}(s)$.

By repeating Fig.~\ref{fig:Sub1to2} three times, to create diamonds
connecting sublattice \#1 to \#2 to \#3 and then back to
\#1, we obtain a structure that links two separated 
diamonds on sublattice \#1, through sites in the other two
sublattices. So
if we restrict ourselves to looking at 
sites in sublattice \#1, 
and the directed percolation structures on that sublattice,
sites that appear disconnected
may be connected by these sites in
sublattices \#2 and \#3. The structure
obtained by repeating Fig.~\ref{fig:Sub1to2} three
times
has 24 sites in sublattices \#2 and
\#3, each of which is occupied with probability $\rho$, 
and has 24
sites in sublattice \#1.
The sites in sublattice \#1 in 
this repeated structure 
map onto the occupied sites in Fig.~\ref{fig:JumpPerc}.
So with $s=\rho^{24}$, $s$ gives the probability of 
having appropriate ``hidden'' occupied sites in sublattices
\#2 and \#3 that connect and make stable the two 
hollow squares. Infinite chains in the jumping directed
percolation model are infinite stable clusters in the knights
model, and thus
$\rho_c^{\rm knights} \leq \rho_c^{\rm Jump}(p^{24}) \leq
\rho_c^{\rm DP}$.

However, since jumping directed percolation is just 
directed percolation with an extra connection process,
it is reasonable to expect that
$\rho_c^{\rm Jump}(s)<\rho_c^{\rm DP}$ for all $s>0$, implying
$\rho_c^{\rm knights}<\rho_c^{\rm DP}$. While this argument is
not mathematically rigorous, it is strongly suggestive, particularly
when we recall previous results on 
enhancements in percolation by
Aizenman and Grimmett~\cite{Enhancement}.
Their work showed that if percolation on a lattice
was ``enhanced'' by adding, for specified subconfigurations,
extra connections or occupations 
with probability $s$, this would
{\it strictly} decrease the critical probability, for 
any $s>0$, so long as the enhancement was essential.
Essential enhancements were defined as those such that
a single enhancement could create a doubly-infinite
path where none existed before.
See Ref.~\cite{Enhancement} for a more rigorous and
precise statement of the results on enhancements,
and Ref.~\cite{PercolationBook} for a general
discussion of enhancements.
The results of Ref.~\cite{Enhancement}
were obtained for undirected percolation, so 
are not directly relevant for
the jumping directed percolation model
considered here,
but they are analogous enough to strongly suggest that
$\rho_c^{\rm Jump}(s)<\rho_c^{\rm DP}$ for all $s>0$.
It is difficult to see how such adding such a new route for
paths to infinity could leave the critical probability 
completely unchanged.


\section{property B}
\label{sec:The implications of Property B}

For models satisfying property A, we have 
$\rho_c=\rho_c^{\rm DP}$,
but we still need to check that the TBF proof that the
transition is glassy (discontinuous with a diverging
crossover length) is valid.

\begin{figure}[tb]
\epsfig{figure=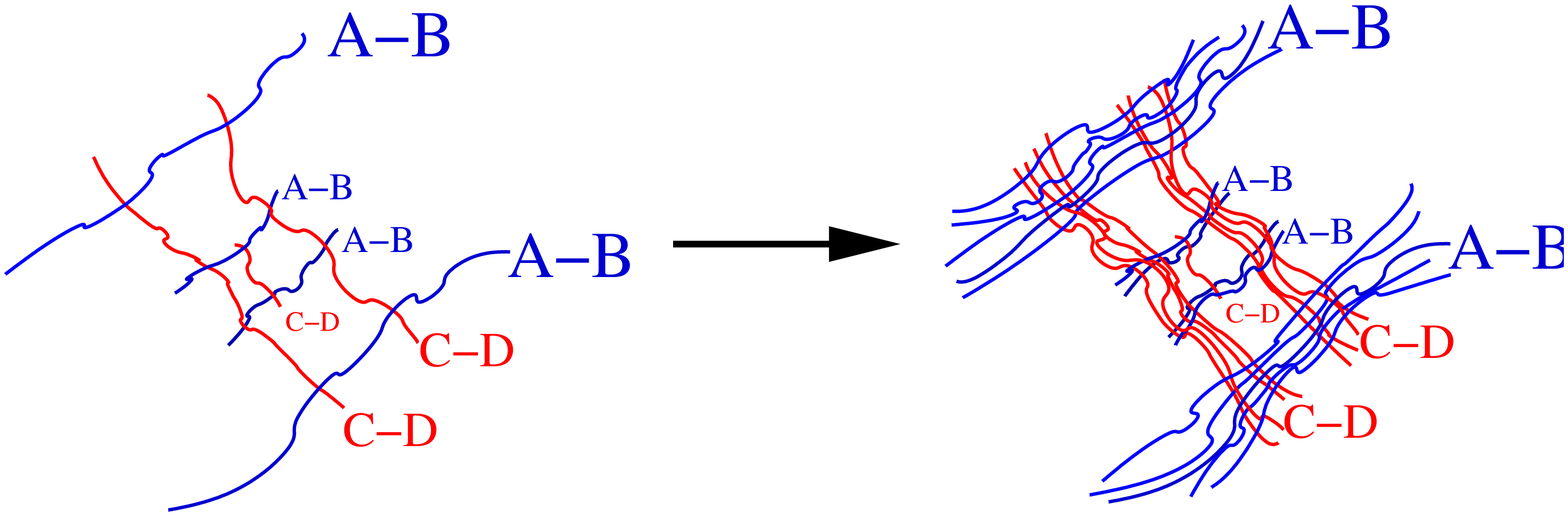,width=3.0in}
\caption{The modification in the TBF discontinuity structure
for models failing the ``No Perpendicular Crossing'' 
rule (Property B)}
\label{fig:TwoFrozenStructures}
\end{figure}

The TBF proof 
of a glassy transition
implicitly assumes that the knights model has property
B. For example, to show discontinuity, they construct a
configuration that has a finite density at 
$\rho_c^{\rm DP}$---see Fig. 2b of~\cite{TBF}.
This figure, and others like it, implicitly assume
property B, because they are based on drawing overlapping
rectangles in independent directions, and assuming that if
paths in these intersecting rectangles cross, they must
stabilize each other (form a T-junction).
The resulting frozen structure is shown on the left 
side of figure~\ref{fig:TwoFrozenStructures}.
However, if a model lacks property B, 
the paths can cross
without stabilizing each other.

The knights model does not satisfy Property A, so 
whether or not it satisfies Property B
is a moot point. 
But what about the sandwich model, which satisfies
property A, but not property B? The TBF proof as it stands is not
immediately valid in these cases. 
Nevertheless, it turns out that for such models, 
the TBF proof can be made to work by a modification of their
structures.

The basic idea of the modification is as follows.
The TBF proof of a glassy transition is based on
drawing 
structures consisting of sets of overlapping
rectangles, showing that there is a 
sufficiently high
probability 
(finite for the proof of discontinuity, and approaching 1
for
the proof of diverging crossover length)
that each rectangle has a spanning path in the
desired direction, and then using property B to conclude
that the intersecting paths form a frozen cluster. 
For models that lack property B, we use the 
same figures as in the TBF proof (e.g. Figs 2a and 2b of
Ref.~\cite{TBF}), but pick the rectangle sizes large enough
that there is a high probability that each rectangle has
{\it multiple} spanning paths
($\mathcal{O}(L^{1-z})$ for a rectangle of $\mathcal{O}(L)$),
each occurring in a
disjoint parallel subrectangle.
Then in each place
where the TBF proof assumes a T-junction based on
property B, we will have a northeast (northwest) path
crossing many northwest (northeast) paths. 
The probability that no
T-junction occurs turns out to decay exponentially with the 
number of northwest (northeast) paths.
In figure~\ref{fig:TwoFrozenStructures} 
we show how the discontinuity structure of
the TBF proof (from Fig. 2a of Ref.~\cite{TBF})
is modified by this procedure.

To implement these ideas, we need to modify
Proposition 5.1 of Ref.~\cite{TB1},
which says that sufficiently large rectangles of size
$aL\times L$ are very likely at the critical point to have
chains connecting the sides of length $aL$:

\bigskip

\noindent {\bf Proposition 5.1} (From Ref.~\cite{TB1}.):\\
For $L\to\infty$ there
exists $c>0$ s.t. \\
$\mu_{L,aL}^{\rho_c} 
(\neg\exists\  
{\rm northeast\  occupied\  cluster\ connecting\
the}\\
{\rm\qquad\qquad sides\ of\ length\ } aL) \\
{\rm \qquad} \leq \exp(-cL^{1-z})$

\bigskip

\noindent Here $z$ is the the dynamical exponent,
 which has been numerically found to be approximately 0.63 in two dimensions~\cite{numericalDP}.
The TBF proof of this Proposition assumes a conjecture that
there is a well-defined
$z<1$ (Conjecture 3.1 of Ref.~\cite{TB1}).
We replace their Proposition 5.1 with the following
proposition,
which instead states that we are likely to have
$\mathcal{O}(L^{1-z})$ connecting chains in disjoint
parallel subrectangles:

\bigskip

\noindent {\bf Proposition 100}:\\
For $L\to\infty$ there exists
$c>0$ and $r>0$ s.t. \\
$\mu_{L,aL}^{\rho_c}
(\neg\exists\ \lfloor rL^{1-z} \rfloor
{\rm \ disjoint\ northeast\ occupied\ clusters,}\\
{\rm\qquad\qquad
occuring\  in\  disjoint\  parallel\ subrectangles,}\\
{\rm \qquad\qquad
connecting\ the\ sides\ of\ length\ } aL) \\
{\rm \qquad}\leq \exp(-cL^{1-z})$

\bigskip

\begin{proof}
We divide the box of size $L$ by $aL$ into
$a L^{1-z}$ parallel disjoint subrectangles, each
of size $L\times L^z$. Assuming
the conjecture of anisotropic 
scaling in directed percolation
(conjecture 3.1 of Ref.~\cite{TB1}),
each subrectangle has a
probability $q>0$ of having a path connecting the
two sizes of length $L^z$, contained within that
subrectangle.
The expected number of crossings is $qa L^{1-z}$, and
for any $r<qa$
the probability of having
less than $rL^{1-z}$ crossings decays
exponentially in $L^{1-z}$.
\end{proof}

The TBF proof of discontinuity shows that for certain 
structures of rectangles,
there is a nonzero probability that each 
rectangle has a suitable ``event,'' and that,
assuming property B, the existence of each
event (a rectangle crossing) results in a stable structure at
the critical point.
Now, with our modified proposition, each ``event''
is the presence of
multiple crossings (in disjoint parallel subrectangles) in
each rectangle, rather than single crossings.

Without property B, this does not guarantee a frozen
cluster.
However,  we see in appendix~\ref{sec:Independence}
that when a northeast
(northwest) path crosses $n$ northwest (northeast) paths,
in $n$ disjoint subrectangles, the
probability of not forming a single
T-junction decays exponentially in $n$.
This is physically obvious, since for large subrectangles,
the probability of each T-junction in each subrectangle is
essentially independent; however, since the probabilities
are not truly independent,
more work is needed to make this rigorous.
The details of the proof
are relegated to
appendix~\ref{sec:Independence}.
More generally, the arguments in
appendix~\ref{sec:Independence}
show that
we can 
treat the probabilities of T-junctions in different
subrectangles as independent, when establishing
an upper bound on the probability. We will use this
throughout this section to multiply such
probabilities as if they were independent.

Assuming property B, the TBF proof shows that the probability of having suitable
events in each rectangle is nonzero at the critical point.
We now need to show that, even without property B, this
results in a finite probability of an appropriate set of
T-junctions.
In the TBF discontinuity structure, shown in
Fig. 6 of Ref.~\cite{TB1}, 
the rectangles are labelled
$\mathcal{R}_i^{1,2}$, $i\geq 1$.
Each rectangle 
$\mathcal{R}_i^{1,2}$ is twice as large as the rectangles
$\mathcal{R}_{i-1}^{1,2}$. So the ``suitable events'' of 
Proposition 100 give
$\mathcal{R}_i^{1,2}$ at least
$k 2^{i(1-z)}$ crossings parallel to its long
direction, where $k$ is some positive constant.
The arguments in 
appendix~\ref{sec:Independence}
show that this results in a T-junction
with
probability
$(1-(1-r)^{k2^{i(1-z)}})$,
for some positive $r$ and $k$.

Starting at the origin, the probability of forming
appropriate T-junctions off to infinity can then be seen to
be

\begin{equation}
(1-(1-r_1)^{k2^{1-z}})^2
\prod_{i=2}^{\infty} (1-(1-r_2)^{k2^{i(1-z)}})^2 \quad ,
\label{eq:SecondS}
\end{equation}

\noindent for some positive $r_1$, $r_2$, and $k$.
This product
converges to a positive number, so the
transition is proven to be discontinuous (subject to 
assumption of the
well-tested conjecture of an anisotropic critical exponent
in directed percolation).

The proof of the diverging crossover length can be made to
avoid the assumption of property B by a similar modification
of the TBF structures.
Again, we begin by repeating the TBF structures, with the
set of parallel rectangles in figure 2a of 
Ref.~\cite{TBF}. In that picture, if every rectangle has a
spanning path, {\it and} the paths all intersect, there will
be a spanning frozen cluster. TBF consider the case where
each rectangle has
sides of order the directed percolation parallel correlation
length, $\xi_{\mid\mid}$. They then 
show that $c_3$ and $c_4$ can be chosen such
that if the system size is 
$L<c_4 \xi_{\mid\mid} \exp [ c_3 \xi_\ell (p)^{1-z} ]$,
the probability that each rectangle is occupied by a
spanning cluster approaches $1$ as $L\to\infty$,
$\rho\to\rho_c^-$.
If property B were to hold, this would result in T-junctions
that would create a frozen structure, and show that the
crossover length diverges as $\rho\to\rho_c^-$.
We no longer have property B; but instead, by
replacing Proposition 5.1 with Proposition 100, 
we can choose the rectangle sizes such that 
each rectangle is occupied by ``many'' spanning clusters
(with ``many'' defined by Proposition 100).
Given this, we can start at an arbitrary rectangle, and then
work our way out, looking for T-junctions to create a
spanning frozen structure. We will only fail to create a
frozen structure if at some point we reach intersecting
rectangles where one spanning path in one rectangle crosses
many spanning paths in the other rectangle, but without
creating a T-junction. By the arguments in appendix A, the
probability of this occurring decays exponentially in
$cL^{1-z}$ for some $c>0$. So even with 
$\mathcal{O}(L/\xi_{\mid\mid})^2$ intersections, the
probability that we ever fail in this process goes to 0 as
$L\to\infty$ and $\rho\to\rho_c^-$, and we are essentially
guaranteed a frozen structure. This shows that the crossover
length diverges as we approach the critical point, with the
same lower bound that TBF found.


\section{Pinwheel model and 8-spiral model}
\label{sec:Pinwheel}

\begin{figure}[tb]
\epsfig{figure=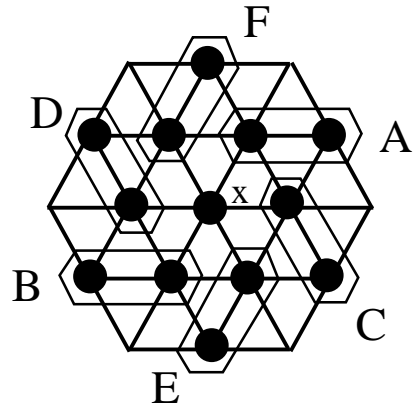,width=2.1in}
\caption{Neighboring relations for the pinwheel model.}
\label{fig:Pinwheel}
\end{figure}

\begin{figure}[tb]
\epsfig{figure=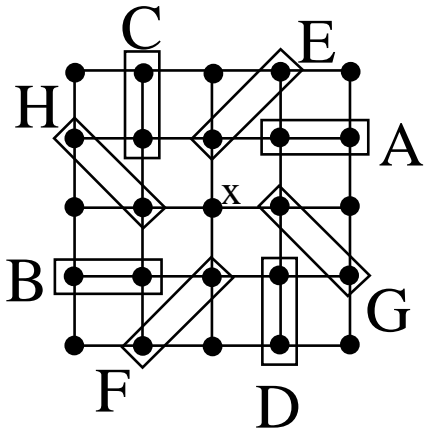,width=2.2in}
\caption{Neighboring relations for the 8-spiral model in which there
are four processes by which a site can be made stable.
For a site to be stable it must have neighbors in A and B,
or C and D, or E and F, or G and H.}
\label{fig:PinwheelPlus}
\end{figure}

The models we have discussed so far have two possible
ways in which a site can be stable, and varying 
neighboring relations for the culling relations. 
However, we can also
define generalizations in which there are three or more ways
in which a site can be stabilized.
For example, for the Pinwheel model, shown
in figure~\ref{fig:Pinwheel},
the condition for a site to be stable is
that it
(1) have neighbors in both the sets A and B, or
(2) have neighbors in both the sets C and D, or
(3) have neighbors in both the sets E and F.
This gives a site three possible directions for stabilizing
chains. Similarly, figure~\ref{fig:PinwheelPlus} shows
a model in which there are four possible directions for a
site to be stable, such that there is an extra ``or'': for the sets G and H. 
We denote this model the 8-spiral model.

Despite the extra ways in which sites can be made stable,
the TBF proof of a glassy transition is still valid, because
Property A holds for both the pinwheel model and the 8-spiral model. 
That is, in both of these models if two
A-B chains (or two C-D chains, or two E-F chains,
or two G-H chains)
cross, they must have sites in common.
This turns out to be sufficient to show that there will be a
void that grows forever in the infinite system limit.

For the sandwich and spiral models, we needed to show that a
stable site at the corner of a diamond-shaped
void has to be part of a directed
percolation-like chain (DP-like chain)
of order the size of the void. For
the pinwheel model, we consider hexagonal voids, and show
that stable sites at the corners of these voids must be
``associated'' with a long DP-like chain,
where ``associated'' will be defined by the construction
below. Then, just as for the sandwich and spiral models, 
for $\rho<\rho_c$, 
long DP-like chains are exponentially suppressed, giving 
voids a finite probability to grow forever, 
showing that the infinite system is empty for
$\rho<\rho_c$.

\begin{figure}[tb]
\epsfig{figure=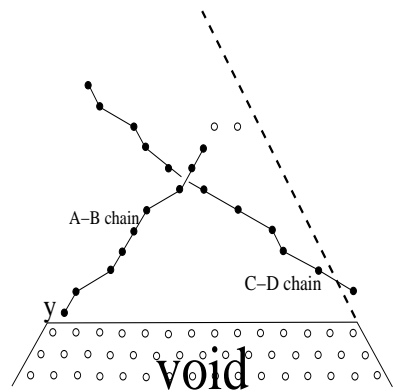,width=2in,height=2in}
\caption{The top of a hexagonal void in the Pinwheel model,
and examples of the A-B and C-D chains discussed in the
text.}
\label{fig:Pinwheel.void}
\end{figure}

Consider the site $y$ at the corner of the hexagonal void in
figure~\ref{fig:Pinwheel.void}. Look at all A-B chains 
coming out of $y$, and pick the lowest possible
chain; in other words, look for successive A neighbors, and
if a site has two possible A neighbors, pick the lower one.
If that chain reaches the dashed line in 
figure~\ref{fig:Pinwheel.void}, we have a
DP-like chain of order the size of the void, and 
are done. Otherwise, this A-B chain must terminate either in
a C-D chain, or an E-F chain. If it terminates in an E-F
chain, that chain must terminate in a C-D chain, which must
then cross the original A-B chain. (Note that the E-F chain
cannot terminate in an A-B chain, since by the ``No Parallel
Crossing'' rule, the new A-B chain would intersect the first
A-B chain, and contradict our assumption that we chose the
lowest A-B chain coming out of the site $y$.)
So if the A-B chain coming out of $y$ does not reach the
dashed line, it must either terminate in a C-D
chain or cross a C-D chain(the latter case is shown in
figure~\ref{fig:Pinwheel.void}). 
Pick the lowest of all the C-D chains that cross
or intersect our
A-B chain. 
This C-D chain must reach the dashed line,
using the same logic as before (if it terminated in an A-B
chain, that would intersect the first A-B chain, and
contradict the assumption that we chose the lowest A-B chain
coming out of $y$; while if it terminated in an E-F chain,
that E-F chain would have to turn into either an A-B or C-D
chain before reaching the void, again resulting in a
contradiction.)
Since the A-B and
C-D chains that we have constructed cross, and together
include both $y$ and the dashed line, at least one of the
chains must be of order the size of the void.  A similar construction 
can also be used for the 8-spiral model using a diamond-shaped void.

Having established where the critical point is, we follow TBF, and
as before
consider an infinite sequence of two types of rectangular 
regions, $\mathcal{R}_{1,2}$ 
(see Fig.~\ref{fig:TwoFrozenStructures}),
one of which contains AB paths,
and the other of which contains EF paths.
(Any pair of types of paths---AB/CD, AB/EF, or CD/EF---are
permissible, as are any of the 6 possible pairs for the
8-spiral model.) Again, the sequence is 
constructed such that the AB paths and the EF paths are mutually 
intersecting with a frozen backbone that contains the
origin, and 
the TBF proof simply carries through with the additional modification 
we have introduced for models that do not obey Property B. 
All in all, since having more neighboring
relations gives more ways for an occupied site to be stable (without 
depressing the DP critical point), the TBF constructions of discontinuous 
percolation structures at the critical point simply carry
through.
Just as for the sandwich 
model,
the TBF proof
needs to be modified to deal with the lack of Property B.
One needs 
at least two intersecting rectangular regions in which the probability 
for two ``transverse'' blocking directions each undergoing a directed 
percolation transition independently is nonzero. 

\section{Models in higher dimensions}

Consideration of the ``No Parallel Crossing'' rule
shows that
for higher-dimensional generalizations of the 
knights model, 
the TBF proof cannot be generalized in a straightforward
manner to show provably glassy transitions.
We will show that the``No Parallel Crossing'' rule never
holds, so the
critical point is always depressed below
that of directed percolation.

To be specific, suppose that 
in three dimensions we have disjoint neighboring sets 
$A$, $B$, $C$, and $D$, generalizing the
northeast, southwest, northwest, and southeast
sets of the knights model (see figure~\ref{fig:knights}).
Each set should consist of three linearly independent
vectors, and the sets $A$ and $B$ should be opposite
of each other, as
should the sets $C$ and $D$. 
Then, just as in the knights model, the culling rule is 
that a stable site should have occupied neighbors in 
both the sets $A$ and $B$, or in both the sets $C$ and $D$.
Then, just as for the sandwich and spiral models, 
there are two directed percolation processes by which a site
can be made stable ($A-B$ chains and $C-D$ chains),
and one might think that for an
appropriate set of neighboring relations the TBF proof could
be used to show a glassy transition.

However, it turns out that because these models 
never satisfy
three-dimensional generalizations of 
the ``No Parallel Crossing'' rule,
the critical point is always
depressed below that of three-dimensional directed
percolation, and the TBF proof cannot be 
directly generalized
for these models. 

Property A says
that two chains running in similar directions cannot cross
without having sites in common.
For certain two-dimensional models, such
as the sandwich and spiral models, these property can be
required by the topology and neighboring relations.
However, in three dimensions, the topology always makes it
easy for two directed chains to miss each other, and so
no three-dimensional generalization of property A
can be satisfied, regardless of the neighboring relations. Furthermore, 
if the two chains miss each other, then the buttressing of each type 
of chain does not occur and the resulting transition may be continuous.

This rough argument can be formalized by showing that
three-dimensional generalizations of the knights model
always have substructures such as the one shown
in figure~\ref{fig:Sub1to2}.
That is, it is always possible to find substructures
that have no long $A-B$ chains connecting their ends, but
which can join up two $A-B$ chains and stabilize their ends.
Then, by the arguments 
in section~\ref{sec:critical.point}, we should expect these
to depress the critical point below that of
three-dimensional
directed percolation.

Specializing to three dimensions for convenience,
let ${\bf A}=\{\vec{a}_1,\vec{a}_2,\vec{a}_3\}$ 
consist of three linearly independent 3-vectors, and
${\bf C}=\{\vec{c}_1,\vec{c}_2,\vec{c}_3\}$ consist 
of three linearly independent 
3-vectors with ${\bf A}\ne {\bf C}$. Also, define  ${\bf B}=\{-\vec{a}_1,-\vec{a}_2,-\vec{a}_3\}$ and  ${\bf D}=\{-\vec{c}_1,-\vec{c}_2,-\vec{c}_3\}$. 
For an occupied site $\vec{r}$ to be stable, it must have either (1) occupied 
neighbors from both $\vec{r}+{\bf A}$ and $\vec{r}+{\bf B}$ or (2) 
occupied neighbors from both $\vec{r}+{\bf C}$ and $\vec{r}+{\bf D}$.
The first condition we denote the A-B condition, the second, the C-D condition. 

If we only enforced the A-B condition, then we would just have 
three dimensional directed percolation (modulo finite clusters).  
However, for the model defined above, there is an extra way
to be stable,
resulting in 
$\rho_c<\rho_c^{DP}$.
Consider a {\em finite} structure 
with the following properties:  (1) all occupied sites are stable under 
the culling rules except occupied sites $\vec{r}_i$ and $\vec{r}_f$, (2) 
$\vec{r}_i$ has an occupied neighbor in $\vec{r}+{\bf B}$ and $\vec{r}_f$ has 
an occupied neighbor in $\vec{r}+{\bf A}$, and (3) there is no AB path 
connecting $\vec{r}_i$ and $\vec{r}_f$, but the structure is stable 
because there exists a path where at least one occupied site is stable 
under the C-D condition.  We relegate to appendix B the proof of the 
existence of such a finite structure.  

With the finite structures defined above, some occupied sites that were 
unstable under the A-B condition now become stable.  Slightly below
$\rho_c^{DP}$, the system is about to percolate using the A-B condition alone.
The substructures act as 
extra, local bonds, joining up long A-B paths and pushing
the system above the critical 
point, just as in the knight model.  Again, arguments 
similar to these
have been made rigorous 
by Aizenmann and Lebowitz
in the case of undirected percolation,
and perhaps can be made rigorous in the 
case of directed (oriented) percolation.

\section{Discussion}

The discovery of a two-dimensional percolation transition, where the sudden emergence of
a  discontinuous backbone 
coincides with a crossover length diverging faster than a power law, 
is recent, and of great interest for glassy systems, 
jamming systems, and phase transitions in general.  Unusual transitions 
have been found previously in mean field systems of a slightly different nature, but not in finite 
dimensions.  And while the finite-dimensional transition is discontinuous, it is not driven 
by nucleation, as with ordinary discontinuous transitions, 
but instead by a scaffolding of many tenuous directed 
percolation paths occurring simultaneously to form a bulky structure. 
We have shown that Property A
is required for the proof that 
$\rho_c=\rho_c^{DP}$, but Property
B is not.
All that is needed to prove that the transition is 
discontinuous (once $\rho_c=\rho_c^{DP}$ is established) is 
a finite probability for two transverse
percolating structures to intersect, to prevent each other
from  being culled.
Therefore, one can construct other models, such as the 
pinwheel and 8-spiral models, 
that exhibit a similar transition in two dimensions. 
The phenomena is not as specific as might seem at first glance.  
However, such a 
buttressing mechanism in dimensions higher than two is more difficult 
because it is more difficult for percolating {\em paths} to intersect 
and form a buttressing, bulky structure. 

Models like the knights model, where $\rho_c$ is most likely 
less than $\rho_c^{DP}$, provide physicists, mathematicians, and 
computer scientists with a motivation to study new models of 
correlated
percolation---models that are not isomorphic to directed percolation, 
but quite possibly in the same universality class.  Once this avenue is pursued further, one can  
then easily extend the class of models for which a finite-temperature transition can be 
rigorously shown. To begin, it would be interesting to consider 
a directed percolation model in two dimensions where the number of 
nearest neighbors is greater than two.  For example, if the number of 
nearest neighbors was increased to four, would the percolation 
transition still be in the same universality class as directed percolation? 
If so, as is presumably the case,
then one could construct a jamming percolation 
model with sets larger than two sites.  These jamming percolation 
models  
would then be isomorphic to the next-neighbor directed percolation models and then
one could use results from directed percolation to prove a   
percolation transition.

There exists another class of correlated 
percolation models called force-balance 
percolation.  The first model in this class was defined in Ref.~\cite{fbperc1}.  Other
force-balance percolation models are currently being constructed and studied ~\cite{fbperc2}.
The force-balance percolation models differ from the jamming percolation 
models in that (1) the sets, such as A, B, etc., are overlapping and (2) the
``or'' between pairs of sets is changed to ``and''.  Given these differences, the 
methods of proof
used here cannot be easily applied.
Numerical results indicate that the transition is discontinuous with a 
nontrivial ``correlation length'' exponent, indicating that the transition
may not the garden-variety discontinuous transition.  The
force-balance models are perhaps less artificial in that they mimic
force-balance by requiring that an occupied site (i.e. a
particle)
have occupied neighbors to its left and right, as well as
its
top and bottom, in order to be stable.
However,
little has been rigorously
proven about them.   To make progress along these lines would be 
useful.

What about the lack of finite stable
clusters in models of jamming percolation? 
Recent numerical work on 
another correlated percolation model, 
$k$-core percolation, 
 with $k=4$ on the four-dimensional hypercubic lattice,
appears to exhibit 
an {\em ordinary}, discontinuous percolation transition driven by nucleation~\cite{rizzo}.  
Finite clusters exist in this model, unlike in the jamming percolation 
models.  Therefore, in the jamming
percolation models, there can be no surface tension between the percolating 
and nonpercolating phases, which is typical of an ordinary discontinuous 
transition.  If finite clusters are allowed in 
a correlated percolation model, one might
guess that the unusual nature of the transition would be 
destroyed.  However, finite clusters, other than individually floating 
particles, do not appear in the jamming transition of
granular particles.
Otherwise, the packing would not be static.  So 
it is unclear whether the
existence of finite 
clusters pertains to the jamming of granular particles.  This is also the 
case for the glass transition. 

If more of an analogy between 
jamming and models of jamming percolation is to be made in
finite dimensions (setting aside the matter 
of the critical dimension of jamming), a model where the fraction of sites participating 
in the infinite cluster increases smaller than linearly just above the 
transition must be found. 
Furthermore, the existence of a jamming percolation model with a 
universal jump in the number of occupied sites at the transition that ``naturally'' emerges as opposed to being externally imposed~\cite{rigidityperc}, is yet another necessary 
quest if a jamming percolation model of jamming is to be found. 

Finally, 
models of correlated percolation, such as the sandwich 
model, tell us that there do indeed exist kinetically constrained models 
of glassy dynamics that exhibit unusual phase transitions in 
finite dimensions. 
Therefore, this avenue of exploration for understanding possible 
finite-temperature glass transitions in finite-dimensions remains 
open.  Since our work helps to clarify which jamming/correlated percolation 
models can be rigorously shown to have an unusual finite-temperature 
glass transition with a particular set of properties, other models exhibiting 
possibly other unusual behaviours can hopefully be more easily developed 
in the near future.


\appendix

\section{Proof that multiple crossings are exponentially
unlikely to avoid creating T-junctions}
\label{sec:Independence}

In this appendix we justify the claim made 
in section~\ref{sec:The implications of Property B}
that given $n$ crossings, 
the probability that no crossing results in a T-junction
decays at least exponentially in $n$.
This would be immediately true if each crossing resulted in
an independent probability of a T-junction.
So what we show is that these crossings, by occurring in
disjoint subrectangles, can be effectively treated
as independent (in establishing an upper bound on the
probability).

\begin{figure}[tb]
\epsfig{figure=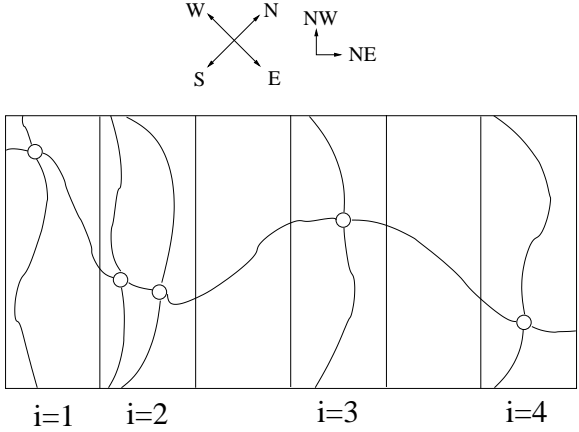,width=3.0in}
\caption{Multiple crossings, resulting in at least one
T-junction with high probability.
Crossings are drawn with open circles to emphasize that
these crossings may or may not result in sites common to the
crossings paths.}
\label{fig:MultipleCrossings}
\end{figure}

The relevant picture is shown for $n=4$ in
Fig.~\ref{fig:MultipleCrossings}.
There are $n$ disjoint rectangles, labeled by $i$,
$1\leq i\leq n$, each of which has at least one
northwest spanning path. 
We call this event $H$, and conditionalize upon the
occurrence of $H$.
Each northwest path must cross the northeast
spanning path, but without Property B,
these crossings do not necessarily result in T-junctions,
where by a T-junction we mean specifically 
a site in common between the
paths that stabilizes the northeast part of the northeast
path.
Let $G_i$ be the event that at least one crossing
in rectangle $i$ forms a T-junction.

\begin{figure}[tb]
\epsfig{figure=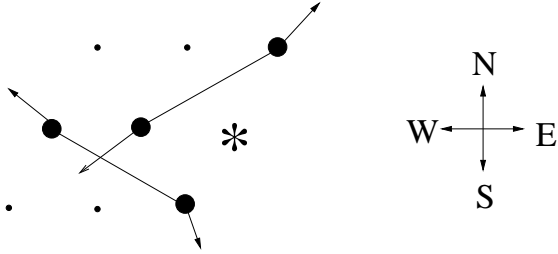,width=2.9in}
\caption{The only configuration in the sandwich model where 
northeast and northwest chains cross, but the 
northeast part of the
northeast chain is not stabilized by a T-junction.
If the site marked with a star 
was occupied, this would create a T-junction.}
\label{fig:MissedCrossing}
\end{figure}

Now for a configuration with $\neg G_i$, look
at the sites in the vicinity of 
the crossing (if there is more than one crossing, we choose
one by an arbitrary ordering of possible crossing
locations). Restricting ourselves first to the 
sandwich model,
Fig.~\ref{fig:MissedCrossing} shows the only way 
that $G_i$ can fail to happen.
The site labeled by a star must be vacant, and if
that site is made occupied, the new configuration is in $G_i$.
So a local change in the vicinity of the 
crossing can always create a T-junction. 
This local change induces a mapping $f_i$ from
the set of states with $\neg G_i$ to 
a subset of the set of states with $G_i$.
The mapping is 
one-to-one onto this subset, and 
for any state $S$ with $\neg G_i$, 
the probability of the configuration $f_i(S)$
is $\rho/(1-\rho)$ times the probability of the
configuration $S$,
where $\rho$ is the site occupation probability.
Thus

\begin{equation}
\mu^\rho (\neg G_i \mid H)\leq \frac{\rho}{1-\rho} \mu^\rho
(G_i\mid H)
\end{equation}

More generally, if we want to consider other variations of
the knights model that lack Property B,
we need only that for any crossing without a
T-junction, some local configuration of changes
in a bounded region around the crossing can create a
T-junction. The induced mapping can be many-to-one,
so long as the ``many'' is bounded (which follows
automatically from the restriction that the 
configuration changes occur in a 
bounded region around the crossing). This will more
generally give

\begin{equation}
\mu^\rho (\neg G_i\mid H)\leq c \mu^\rho (G_i\mid H) \ \ ,
\end{equation}

\noindent for some $0<c<1$.

Since the configuration changes only take place within
a rectangle $i$, they do not affect whether or not we have
$G_j$ for $j\neq i$, and we can write the above inequality
for a state where we specify whether these other 
$G_j$ occur. For example, for $i=3$ we might write

\begin{eqnarray}
\nonumber
\mu^\rho (G_1 \cap \neg G_2 \cap \neg G_3 \cap G_4 \dots
\cap G_n\mid H)  & \leq &
\\
c
\mu^\rho (G_1 \cap \neg G_2 \cap G_3 \cap G_4 \dots
\cap G_n & & \hspace{-0.2in} \mid   H)  \ \ , 
\label{eq:SampleG}
\end{eqnarray}

\noindent with the same $c$ as above. This 
can be intuitively thought of as treating the different
probabilities of forming T-junctions as independent.
Repeatedly using equation~\ref{eq:SampleG}
in different subrectangles, we find that

\begin{equation}
\mu^\rho \left(\bigcap_{i=1}^n \neg G_i \mid H \right) \leq
\frac{1}{(1+c)^n}
\end{equation}

\noindent We are thus exponentially unlikely to have no
T-junctions.

\section{Existence of finite structures in three-dimensional models}

In this appendix we prove the existence of the finite structures discussed 
in the section on jamming percolation in three-dimensions.  
Before giving the formal proof, we sketch the qualitative
idea behind the construction of these structures.
In the structure for the knights
model, shown in
figure~\ref{fig:Sub1to2}, 
there are two parallelograms consisting of $A-B$
(northeast-southwest) chains. The two parallelogragrams are
parallel to each other, and are connected and
stabilized by $C-D$ (northwest-southeast) chains. In 
two dimensions, such a figure can only be constructed
by having some chains cross each other,
and this results in a long $A-B$ chain connecting the
two ends, unless the model violates Property A.
However, in three dimensions, the $C-D$ chains can always
be run in a direction independent of the plane of the $A-B$
parallelograms, so such a substructure can always be formed,
regardless of the neighboring relations. 

We now formalize this argument.
Recalling that $\aone$, $\atwo$, $\athree$, and $\cone$ are four
three-dimensional vectors, with the first three being linearly independent and 
$\cone$ not equal to any of the first three, there must exist $m_1$, $m_2$,
$m_3$, and $n$ such that

\begin{equation}
\sum_{i=1}^3 m_i \vec{a}_i = n \vec{c}_1
\label{eq:lineardependence}
\end{equation}

\noindent Since the $\vec{a}_i$ are linearly independent of
each other, $n\neq 0$. Since $\vec{c}_1\notin {\bf A}$,
at least two of the $m_i$ must be nonzero.

\begin{figure}[tb]
\epsfig{figure=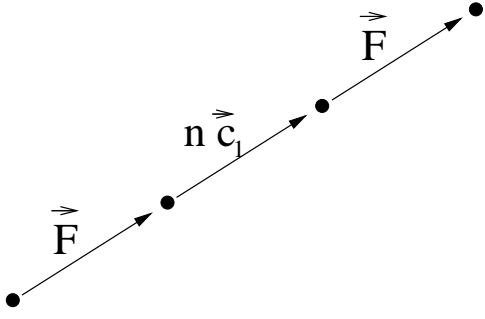,width=2.5in}
\caption{The structure in the case where
$\vec{c}_1=\sum_{i=1}^3 m_i \vec{a}_i$ with all
$m_i>0$.}
\label{fig:simpleline}
\end{figure}

If all of the $m_i$ are positive, then it is easy to make
the desired structure. Let
$\vec{F}\equiv \sum_{i=1}^3 m_i \vec{a}_i = n \vec{c}_1$.
Then make the structure in figure~\ref{fig:simpleline},
where each of the vectors $\vec{F}$ represents an A-B
chain of length $\sum_{i=1}^3 m_i$, and
$n c_i$ represents a C-D chain of length $n$.
This structure has the desired properties.
If all of the $m_i$ are negative, we simply replace
${\bf C}$ with ${\bf D}$ and use the same argument.

We are now left with the most complicated case, where
some $m_i$ are positive, and some $m_i$ are
negative. In this case, we redefine the $m_i$, and rewrite
equation~\ref{eq:lineardependence} as

\begin{equation}
\sum_{i=1}^3 m_i \vec{a}_i = \sum_{i=1}^3 \tilde{m}_i
\vec{a}_i + n \vec{c}_1 \quad ,
\end{equation}

\noindent where $n$, all $m_i$, and all $\tilde{m}_i$, are
positive, and for any $i$, either $m_i$ or $\tilde{m}_i$ is
zero. Then define $\vec{F}\equiv\sum_{i=1}^3 m_i \vec{a}_i$ and
$\vec{G}\equiv\sum_{i=1}^3 \tilde{m}_i \vec{a}_i$.
$\vec{F}$ and $\vec{G}$ are both nonzero and linearly
independent.

\begin{figure}[tb]
\epsfig{figure=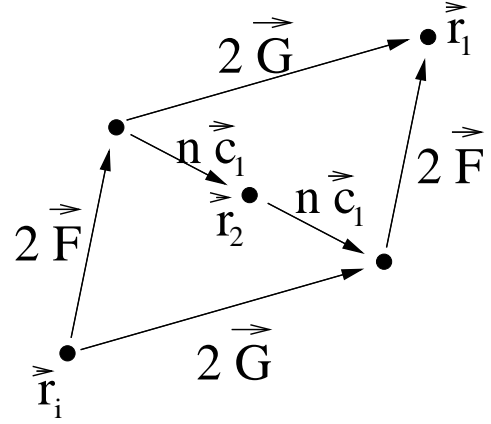,width=2.5in}
\caption{A substructure}
\label{fig:substructure}
\end{figure}

We can now make the structure shown in
figure~\ref{fig:substructure}
The vectors $2\vec{F}$ and $2\vec{G}$ represent A-B
chains of length $2\sum_{i=1}^3 m_i$ and
$2\sum_{i=1}^3 \tilde{m}_i$. In these chains every site is
stable by the A-B condition
except for $\vec{r}_i$ and
$\vec{r}_1$. The vectors $n\vec{c}_1$ are chains of length
$n$ in which every site is stable by the C-D condition.

\begin{figure}[tb]
\epsfig{figure=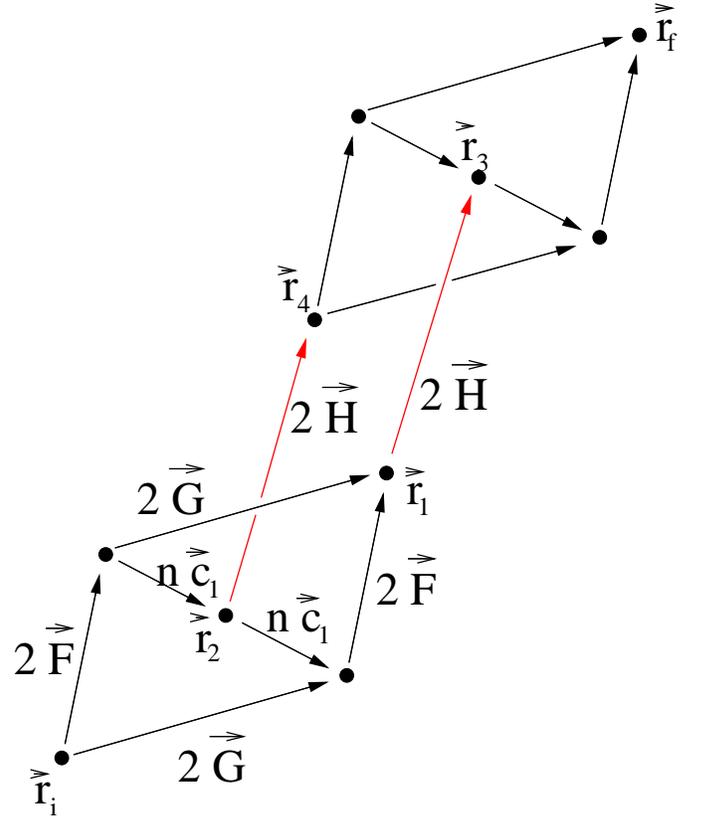,width=3.5in}
\caption{The full structure with the desired properties}
\label{fig:fullstructure}
\end{figure}

Next, since $\{ \vec{a}_1, \vec{a}_2, \vec{a}_3\}$ form a
complete basis for three-dimensional space, and
$\{\vec{F}, \vec{G}\}$ are only two vectors, there
exists a vector $\vec{H}=\sum_{i=1}^3 p_i \vec{a}_i$,
with all $p_i>0$, such that $\vec{H}$ is linearly independent
of $\{\vec{F},\vec{G}\}$.
(Note: this is where the three-dimensional case differs from
the two-dimensional case. For models such as the spiral and
sandwich models, the two vectors $\vec{F}$ and $\vec{G}$
already span the space.)
We can then make a second copy of
figure~\ref{fig:substructure},
displaced from the original by $\vec{H}$,
as shown in figure~\ref{fig:fullstructure}.

We can now check that in figure~\ref{fig:fullstructure}
all sites except the
start and end sites, $\vec{r}_i$ and $\vec{r}_f$ are stable
under the culling rules, and that $\vec{r}_i$ and
$\vec{r}_f$ have neighbors from ${\bf B}$ and ${\bf A}$,
respectively. It remains to check that there is no A-B chain
connecting $\vec{r}_i$ to $\vec{r}_f$ in this structure.
There is no obvious such chain, but depending on the
vectors $\vec{a}_i$ and $\vec{c}_1$, it is possible that
there are some sites in the
$\vec{F}$, $\vec{G}$, and $\vec{H}$ chains by chance are
separated by a vector $\vec{a}_i$, inadvertently
forming an A-B chain between $\vec{r}_i$ and $\vec{r}_f$.
However, if this is the case, we can create new larger
structure, simply by multiplying $n$, all $m_i$, all
$\tilde{m}_i$, and all $p_i$ by the same multiplicative
constant. The structure thus grows larger, while the vectors
$\vec{a}_i$ stay the same, so for a sufficiently large
multiplicative constant, it is impossible for the different
chains in the structure to be adjacent by $\vec{a}_i$
connections. We thus have a structure with the desired
properties.



\begin{thebibliography}{9}
\bibitem{unique} The Kosterlitz-Thouless transition also exhibits 
unusual characteristics. There is a discontinity in the spin wave stiffness, 
or the free energy cost to applying a gradient,   
accompanied with an exponentially diverging correlation length.  However, 
there is no magnetization, or local order parameter, that goes to 
zero at the transition. The jamming percolation models have a simple 
order parameter, the fraction of sites in the infinite cluster.
\end{thebibliography}
\end{document}